\documentclass[3p,times]{elsarticle}

\usepackage{ecrc}

\volume{00}

\firstpage{1}

\journalname{Nuclear Physics A}

\runauth{}

\jid{nupha}

\usepackage{amssymb}

\usepackage[figuresright]{rotating}

\renewcommand{\b}[1]{{\bf #1}} 
\newcommand{\beq}{\begin{equation}}
\newcommand{\beql}[1]{\begin{equation}\label{#1}}
\newcommand{\eeq}{\end{equation}}
\newcommand{\bea}{\begin{eqnarray}}
\newcommand{\eea}{\end{eqnarray}}
\newcommand{\eq}[1]{(\ref{#1})}
\newcommand{\fig}[1]{Fig.~\ref{#1}}

\newcounter{topiccounter}
\renewcommand{\b}[1]{\mathbf{#1}}

\newcommand{\as}{\alpha_s}

\usepackage{epsfig,epsf}

\begin{document}

\begin{frontmatter}



\dochead{}

\title{Rapidity and centrality dependence of azimuthal correlations in high energy d+Au collisions}


\author{Kirill Tuchin}

\address{Department of Physics and Astronomy, Iowa State University, Ames, IA 50011}

\begin{abstract}
We  discuss azimuthal correlations  in $dAu$ collisions at different rapidities and centralities  and argue that experimentally observed depletion of the back-to-back bump can be quantitatively explained by gluon saturation  in the Color Glass Condensate of the Gold nucleus.  
\end{abstract}

\begin{keyword}


\end{keyword}

\end{frontmatter}


\section{ Introduction}
\label{}
In a pioneering paper \cite{Kharzeev:2004bw} it was proposed to study the azimuthal correlations of hadrons produced at large rapidity separation $\Delta y\gg 1$. The idea is that such correlations are mediated by the BFKL Pomeron.  Therefore, unlike hadron production in hard collisions, where there is strong back-to-back correlation at opening azimuthal  angle $\Delta\phi= \pi$, correlations in the CGC should be significantly reduced. It has been suggested in \cite{Marquet:2007vb} that correlations at small $\Delta y$ in the forward direction can also be used to study CGC. Indeed, forward direction corresponds to small $x$ of nucleus where the CGC effects are strongest. They reduce both single and double inclusive hadron production and thus back-to-back correlations are suppressed. 
Unfortunately, there is a technical problem: the relevant scattering amplitudes are well-known in the so-called Multi-Regge-Kinematics  $\Delta y\gg 1$, which  is not applicable  in this case.  One therefore has to rely on phenomenological 
models, which offer descriptions that are analytically accurate only in parts of the interesting kinematic region. There are two such approaches: one that is based on the dipole model \cite{Marquet:2007vb,Albacete:2010pg} and another one that is based on the $k_T$-factorization \cite{Tuchin:2009nf}.

The present calculation, based on  `$k_T$-factorization',  assumes that $2\to n$ process and  the two-point correlation functions of CGC fields  can be factored out. In this approximation,  the $2\to 4$ amplitudes were calculated for an arbitrary $\Delta y$ (quasi multi-Regge kinematics, QMRK) in \cite{LRSS,CCH,CE,Fadin:1997hr} for $gg\to ggq\bar q$ and in \cite{Fadin:1996zv,Leonidov:1999nc,Bartels:2006hg} for $gg\to gggg$ processes. 
  Although generally  $k_T$-factorization fails in the gluon saturation region, there are valid reasons to believe that it provides a \emph{reasonable approximation} of the observed quantities. Indeed, it was proved that $k_T$-factorization provides the exact result for the cross section for single inclusive gluon production in the leading logarithmic approximation (LLA) \eq{sincl} \cite{Kovchegov:2001sc} (though there is  a subtlety in the definition of the unintegrated gluon distribution $\varphi$ \cite{Kovchegov:2001sc,Kharzeev:2003wz}). Although $k_T$-factorization fails for the double-inclusive heavy quark production,  the deviation from the exact results is not large at RHIC energies  \cite{Fujii:2005vj}. At  transverse momenta of produced particles much larger than $Q_s$, $k_T$-factorization rapidly converges to the exact results. There are also numerous indications that  $k_T$-factorization is  phenomenologically reliable (see \cite{Tuchin:2009nf} for examples). 







\section{Correlations at $|y_T-y_A|\lesssim 1$ }\label{sec2}

First, we would like to consider correlations at small rapidity separations. 
Azimuthal correlation function is defined as 
\beq\label{cf}
C(\Delta \phi)=\frac{1}{N_{\mathrm{trig}}}\frac{dN}{d(\Delta\phi)}\,,
\eeq
where $dN/d(\Delta\phi)$ is the number of pairs produced in the given opening angle $\Delta\phi$ and $N_{\mathrm{trig}}$ is the number of trigger particles. 
The number of pairs is given by
\beq\label{npa}
\frac{dN}{d(\Delta\phi)}= 2\pi \int dk_T k_T \int d y_T\,\int  dk_A k_A \int dy_A
\left( \frac{dN_\mathrm{trig}}{d^2k_Tdy_T}  \frac{dN_\mathrm{ass}}{d^2k_Ady_A}
+ \frac{dN_\mathrm{corr}}{d^2k_Tdy_T\,d^2k_Ady_A}  \right)
\eeq
where $\b k_T$ and $y_T$ are the transverse momentum and rapidity  of the trigger particle and   $\b k_A$ and $y_A$ are the transverse momentum and rapidity  of the associate one. We denote $k_T=\sqrt{\b k_T^2}$ etc.\ throughout this paper.
The first term on the r.h.s.\ of \eq{npa} corresponds to gluon production in two different  sub-collisions (i.e.\ at different impact parameters) and therefore gives a constant contribution to the correlation function, whereas the second term on the r.h.s.\  describes production of two particles in the same sub-collision.  The number of the trigger particles is given by 
\beq\label{strig}
N_\mathrm{trig}= 2\pi \int dk_T k_T\int dy_T\, \frac{dN_\mathrm{trig}}{d^2k_Tdy_T}\,.
\eeq
Expression for the single inclusive  gluon cross section is well-known (see e.g.\ \cite{Kovchegov:2001sc}). The corresponding multiplicity reads
 \beql{sincl}
 \frac{dN}{d^2 k\,dy}\,=\,
\frac{2 \, \as\,}{C_F \, S_\bot}\,\frac{1}{ k^2}\,\int\,
d^2 q_1\varphi_D(x_+,q^2_1)\,\varphi_A(x_-,(\b k- \b q_1)^2)\,.
\eeq
In the center-of-mass frame $x_\pm=\frac{k}{\sqrt{s}}\, \exp\{\pm y\}\,.$
Equation \eq{sincl} is derived in multi-Regge kinematics (MRK) $x_\pm\ll 1$. 

The correlated part of  double-inclusive parton multiplicity is given by 
\begin{eqnarray}
\frac{dN_\mathrm{corr}}{d^2 k_T\,dy_T\, d^2 k_A\,dy_A}&=&
\frac{N_c \, \as^2}{\pi^2 \, C_F \, S_\bot}\,
\int\,
\frac{d^2 q_1}{q_1^2}\,\int\,\frac{d^2q_2}{q_2^2}\,\delta^2(\b q_1+\b q_2-\b k_T-\b k_A)\nonumber\\
&&\times\,\varphi_D(x_1,q^2_1)\,\varphi_A(x_2,q^2_2)\,\mathcal{A}(\b q_1,\b q_2, \b k_T, \b k_A,y_T-y_A)\,,\label{nlow}
\end{eqnarray}
where $x_{1,2}=(k_{T}e^{\pm y_T} + k_{A}e^{\pm y_A})/\sqrt{s}$.
The amplitude $\mathcal{A}$ was computed in the quasi-multi-Regge-kinematics (QMRK) in \cite{Fadin:1996zv,Fadin:1997hr,Leonidov:1999nc} and recently re-derived in \cite{Bartels:2006hg} (the $gg\to ggq\bar q $ part was calculated before in \cite{LRSS,CCH,CE}). In QMRK one assumes that $x_1,x_2\ll 1$, but $\Delta y $ is finite. 
Explicit expression for $\mathcal{A}$  can be found in  \cite{Leonidov:1999nc}.

For numerical calculations we need a model for the unintegrated gluon distribution function $\varphi$.  In spirit of the KLN model \cite{Kharzeev:2001yq} we write 
\beq\label{kln}
\varphi(x,q^2)=\frac{1}{2\pi^2}\frac{S_\bot C_F}{\as}\big(1-e^{-Q_s^2/q^2}\big) \, (1-x)^4\,.
\eeq 
where the saturation scale of nucleus is $Q_s^2= A^{1/3}Q_{sp}^2$, with $Q_{sp}^2$ the saturation scale of proton fixed by fits of the DIS data. The coupling constant is fixed at $\as=0.3$. 

It has been pointed out in \cite{Leonidov:1999nc} that due to $1\to 2$ gluon splittings the double-inclusive cross section has a collinear  singularity at $\hat s\to 0$, i.e.\ it is proportional to $[(\Delta y)^2  +(\Delta \phi)^2]^{-1}$. Such singularities are usually cured at higher orders of perturbation theory. Additional contributions to the small angle correlations arise from various soft processes including resonance decays, hadronization, HBT correlations etc. Because the small angle correlations are beyond the focus of the present paper we simply regulate it by imposing a cutoff on the  minimal possible value of the invariant mass $\hat s$. This is done by redefining the amplitude as $\mathcal{A}\to \mathcal {A} \,\hat s/(\mu^2+\hat s)$. For each kinematic region, parameter $\mu$ is fixed in such a way as to reproduce the value of the correlation function in $pp$ collisions  at zero opening angle $\Delta\phi=0$. 

 $k_T$-factorization is known to give results that are in qualitative agreement with a more accurate approaches, but miss the overall normalization. Therefore, in order to correct the overall normalization of the cross sections we multiply the single inclusive cross section \eq{sincl} by a constant $K_1$ and the double-inclusive one \eq{nlow} by a different constant  $K_2$ \cite{Kovchegov:2002nf,Kovchegov:2002cd}. The correlation function $C$ depends on both $K_1$ and $K_2$. However, the difference $C_\Delta=C(\Delta\phi)-C(\Delta\phi_0)$ depends only on the ratio $K_2/K_1$. We choose $\Delta\phi_0$ in such a way  that $C(\Delta\phi_0)$ is the minimum of the correlation function. This is  analogous to the experimental procedure of removing the pedestal \cite{Adams:2003im}. The overall normalization of  the correlation function $K_2/K_1$ -- which is the only essential free parameter of our model -- is fixed to reproduce the height of the correlation function in $pp$ collisions. 

\begin{figure}[ht]
\begin{tabular}{cc}
      \includegraphics[height=4.5cm]{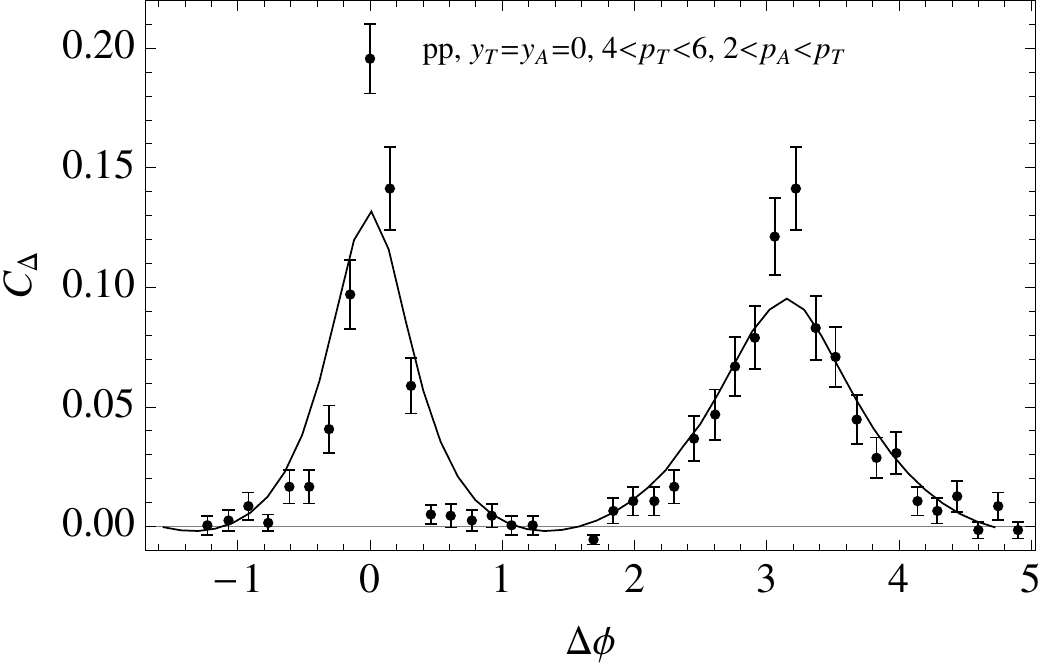} &
      \includegraphics[height=4.5cm]{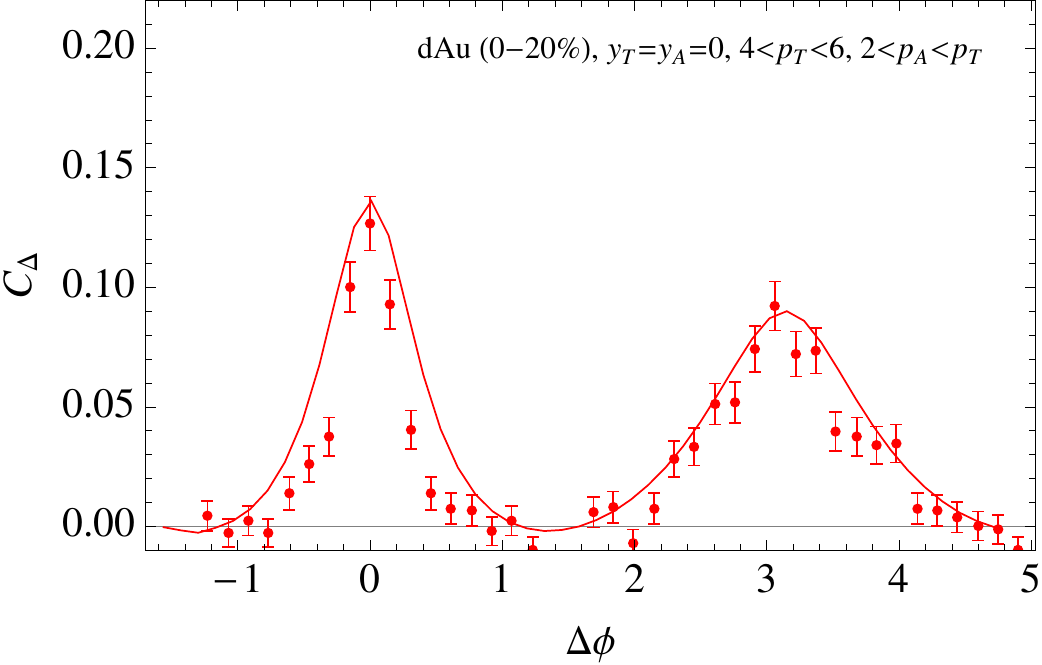} 
  \end{tabular}
  \caption{Correlation function at central rapidity. Kinematic region is  $4<p_T <6$, $2<p_A <p_T$ (all momenta are in GeV), $y_T=3.1$, $y_A=3$. Left (right) panel: minbias $pp$ ($dAu$) collisions. Data from \cite{Adams:2003im}. }
\label{fig:cc}
\end{figure}

\begin{figure}[ht]
\begin{tabular}{cc}
      \includegraphics[height=4.5cm]{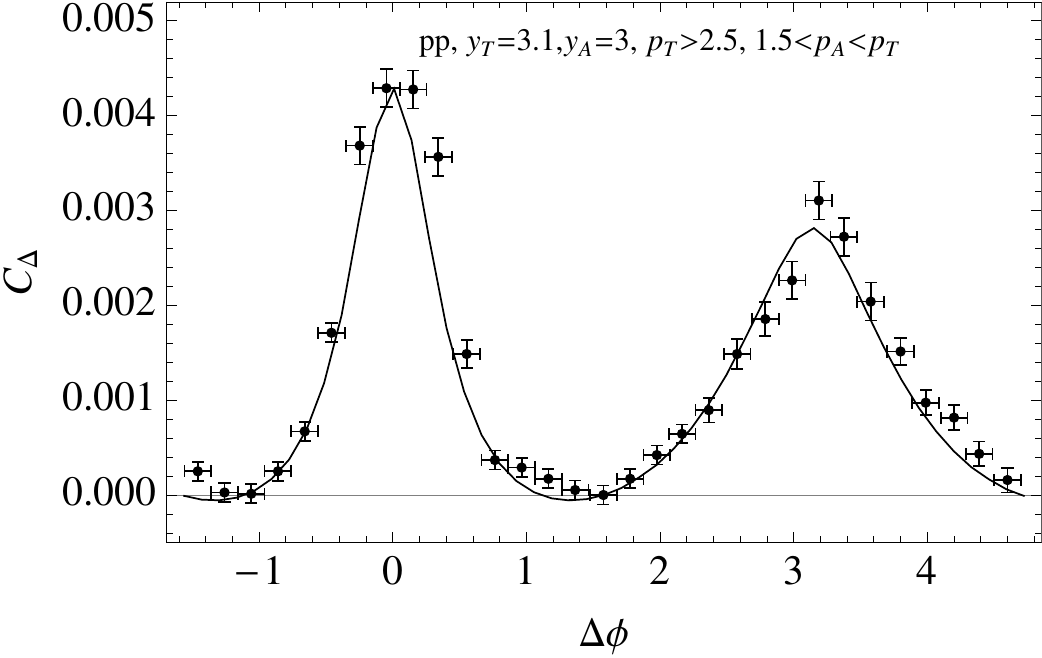} &
      \includegraphics[height=4.5cm]{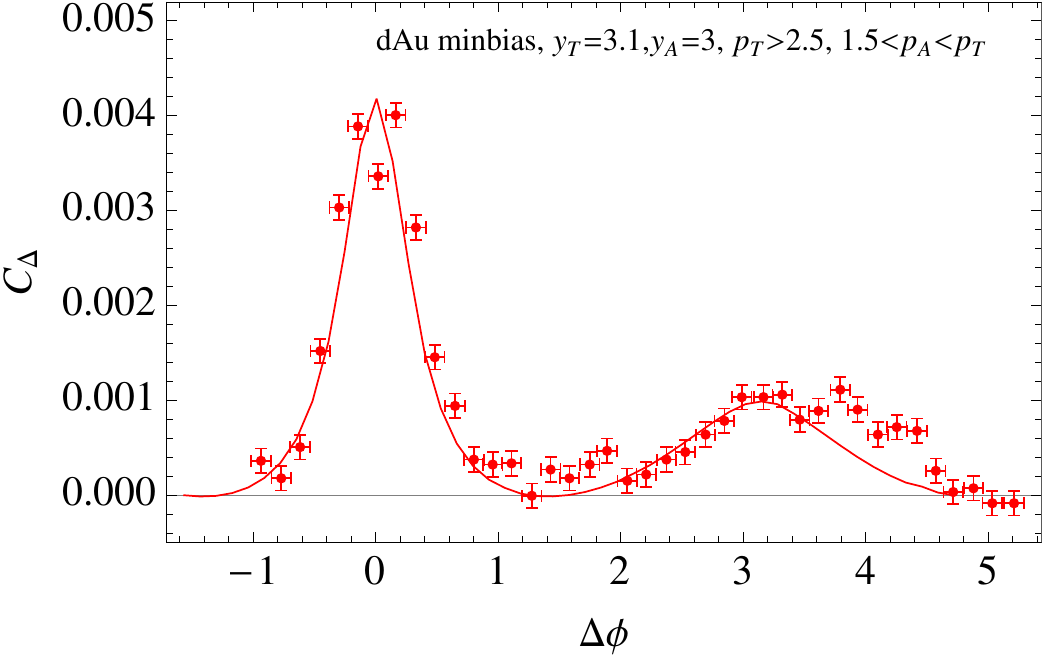} 
  \end{tabular}
  \caption{Correlation function at forward rapidities. Kinematic region is  $p_T>4$, $1.5<p_A <p_T$ (all momenta are in GeV), $y_T=3.1$, $y_A=3$. Left (right) panel:  the minbias $pp$ ($dAu$) collisions. Data from \cite{AGordon}. }
\label{fig:ff1}
\end{figure}

\begin{figure}[ht]
\begin{tabular}{cc}
      \includegraphics[height=4.5cm]{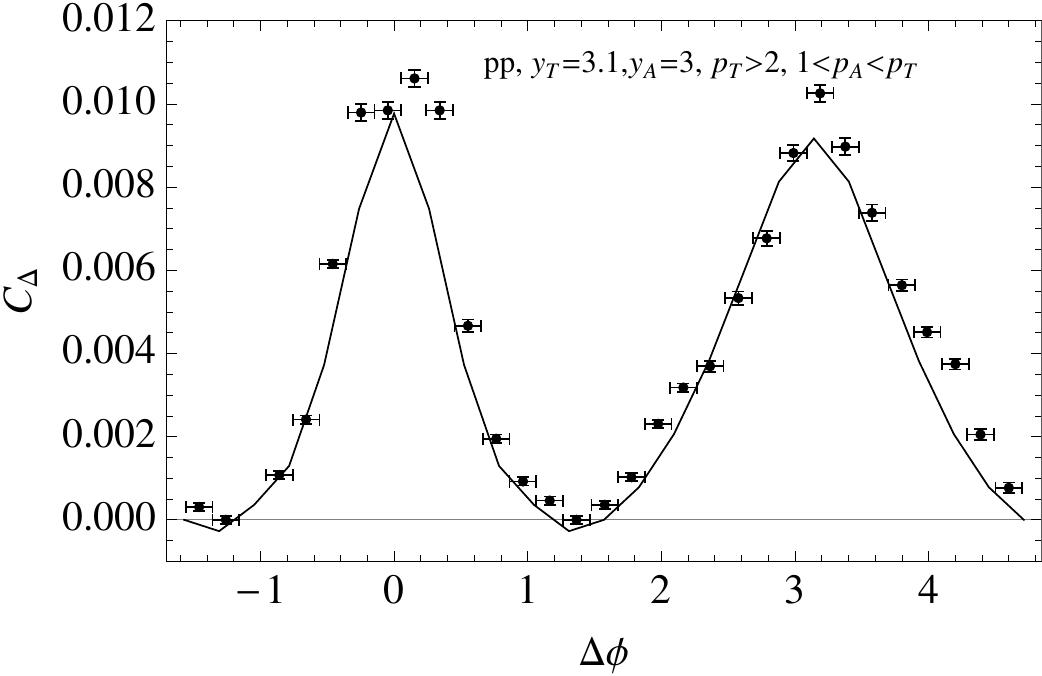} &
      \includegraphics[height=4.5cm]{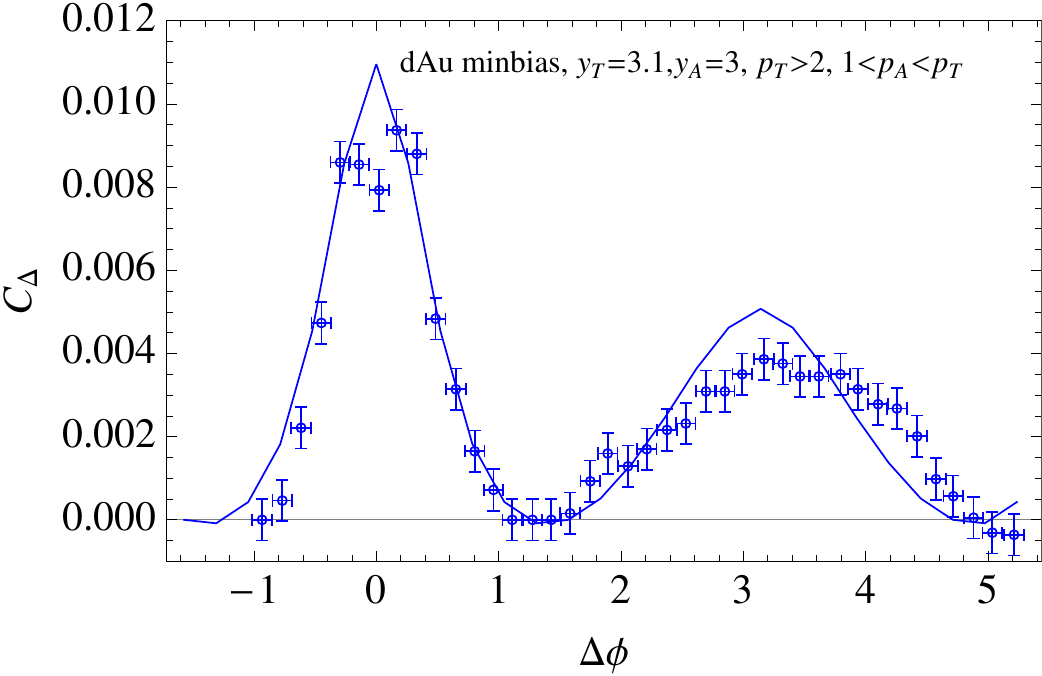} \\
      \includegraphics[height=4.5cm]{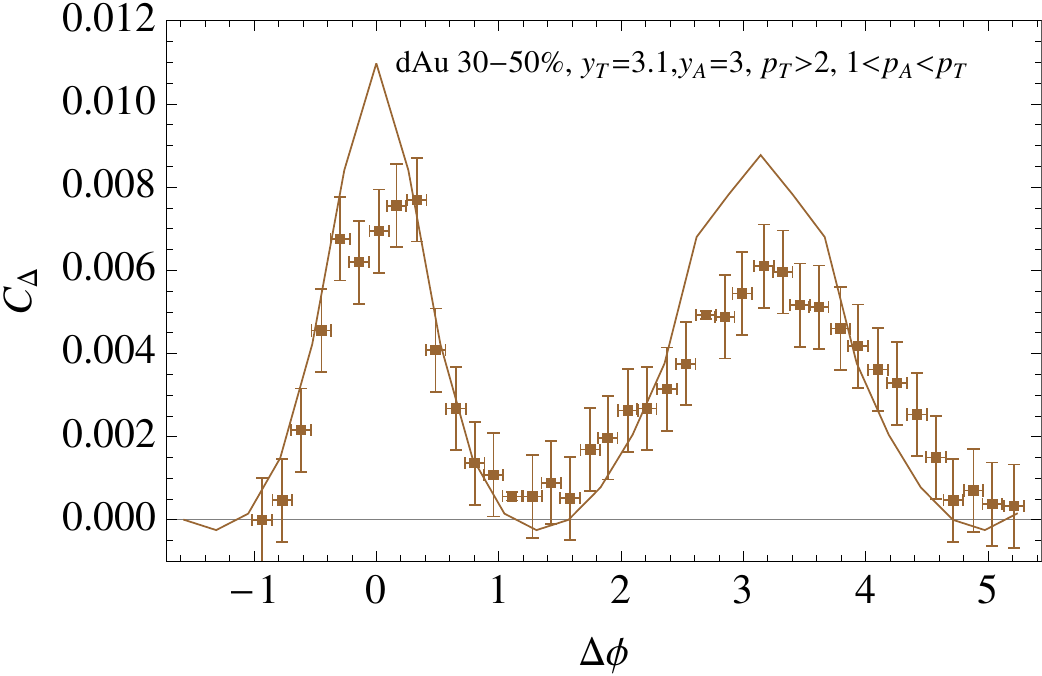}&
      \includegraphics[height=4.5cm]{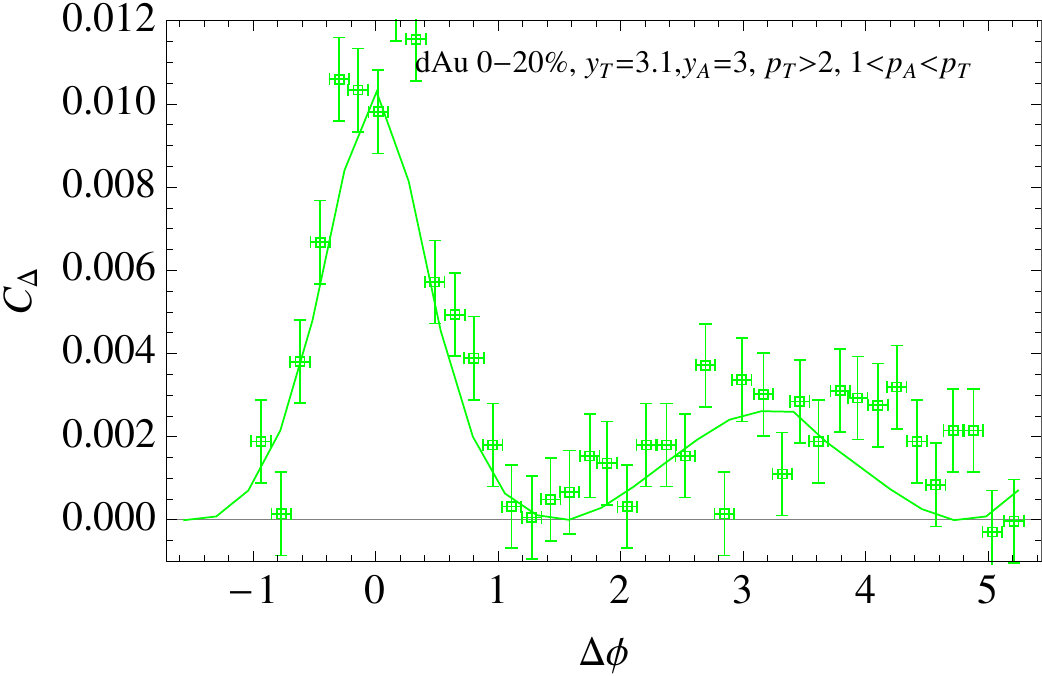}
  \end{tabular}
  \caption{Correlation function at forward rapidities. Kinematic region is  $p_T>2$, $1.5<p_A <p_T$ (all momenta are in GeV), $y_T=3.1$, $y_A=3$. Upper left (right) panel: minbias $pp$  ($dAu$) collisions. Lower left (right) panel: peripheral (central)  $dAu$ collisions. Note: centrality of the theoretical calculation may not coincide with the centrality of the data (the former is not yet known at the time of publication).   Data from \cite{AGordon}.}
\label{fig:ff2}
\end{figure}

The results of the numerical calculations are shown in Fig.~\ref{fig:cc}--\ref{fig:ff2}.  In these figures we observe suppression of the bak-to-back correlation in $dAu$ as compared to the bak-to-back correlation in $pp$, in agreement with the  experimental data. In \fig{fig:ff2} we also see the depletion of the  back-to-back correlation as a function of centrality. Note, that at the time of publication the precise centrality classes of the \emph{data} shown in the lower row of \fig{fig:ff2} were not known.     

In addition to  $gg\to gggg$ and $gg\to ggq\bar q$ processes that we took into account in this section, production of valence quark of deuteron $gq_v\to gq_v gg$ gives a sizable contribution at forward rapidities due to moderate value of $x$ associated with deuteron ($x\approx 0.2$ for $p_T=2$~GeV at $y=3$). Contribution of this process to azimuthal correlations was analyzed in \cite{Marquet:2007vb} in the framework of the dipole model in MRK. However, the corresponding expression  in $k_T$-factorization in QMRK is presently unknown thus preventing us from taking it into account in our calculation.  In-spite of this we believe that the general structure of the correlation function as well as its centrality dependence is not strongly affected by the valence quark contribution. We plan to address this problem elsewhere.

\section*{Acknowledgements}
This work  was supported in part by the U.S. Department of Energy under Grant No.\ DE-FG02-87ER40371.


\begin{thebibliography}{40}



  
  
\bibitem{Kharzeev:2004bw}
  D.~Kharzeev, E.~Levin and L.~McLerran,
  Nucl.\ Phys.\  A {\bf 748}, 627 (2005)
  [arXiv:hep-ph/0403271].


\bibitem{Marquet:2007vb}
  C.~Marquet,
  Nucl.\ Phys.\  A {\bf 796}, 41 (2007)
  [arXiv:0708.0231 [hep-ph]].


  
\bibitem{Albacete:2010pg}
  J.~L.~Albacete and C.~Marquet,
  arXiv:1005.4065 [hep-ph].

\bibitem{Tuchin:2009nf}
  K.~Tuchin,
  Nucl.\ Phys.\  A {\bf 846}, 83 (2010)
  [arXiv:0912.5479 [hep-ph]].

\bibitem{LRSS}
E.~M.~Levin, M.~G.~Ryskin, Y.~M.~Shabelski and A.~G.~Shuvaev,
Sov.\ J.\ Nucl.\ Phys.\  {\bf 53}, 657 (1991)
[Yad.\ Fiz.\  {\bf 53}, 1059 (1991)].

\bibitem{CCH}
S.~Catani, M.~Ciafaloni and F.~Hautmann,
Nucl.\ Phys.\ B {\bf 366}, 135 (1991).

\bibitem{CE} 
J.~C.~Collins and R.~K.~Ellis,
Nucl.\ Phys.\ B {\bf 360}, 3 (1991).

\bibitem{Fadin:1996zv}
  V.~S.~Fadin, M.~I.~Kotsky and L.~N.~Lipatov,
  arXiv:hep-ph/9704267.

\bibitem{Fadin:1997hr}
  V.~S.~Fadin, R.~Fiore, A.~Flachi and M.~I.~Kotsky,
  Phys.\ Lett.\  B {\bf 422}, 287 (1998)
  [arXiv:hep-ph/9711427].

\bibitem{Leonidov:1999nc}
  A.~Leonidov and D.~Ostrovsky,
  Phys.\ Rev.\  D {\bf 62}, 094009 (2000)
  [arXiv:hep-ph/9905496].
  
\bibitem{Bartels:2006hg}
  J.~Bartels, A.~Sabio Vera and F.~Schwennsen,
  JHEP {\bf 0611}, 051 (2006)
  [arXiv:hep-ph/0608154].


\bibitem{Kharzeev:2003wz}
  D.~Kharzeev, Y.~V.~Kovchegov and K.~Tuchin,
  Phys.\ Rev.\  D {\bf 68}, 094013 (2003)
  [arXiv:hep-ph/0307037].


\bibitem{Kovchegov:2001sc}
  Y.~V.~Kovchegov and K.~Tuchin,
  Phys.\ Rev.\  D {\bf 65}, 074026 (2002)
  [arXiv:hep-ph/0111362].

\bibitem{Fujii:2005vj}
  H.~Fujii, F.~Gelis and R.~Venugopalan,
  Phys.\ Rev.\ Lett.\  {\bf 95}, 162002 (2005)
  [arXiv:hep-ph/0504047].
  
\bibitem{Kharzeev:2001yq}
  D.~Kharzeev, E.~Levin and M.~Nardi,
  Phys.\ Rev.\  C {\bf 71}, 054903 (2005)
  [arXiv:hep-ph/0111315].
  
\bibitem{Kovchegov:2002nf}
  Y.~V.~Kovchegov and K.~L.~Tuchin,
  Nucl.\ Phys.\  A {\bf 708}, 413 (2002)
  [arXiv:hep-ph/0203213].

\bibitem{Kovchegov:2002cd}
  Y.~V.~Kovchegov and K.~L.~Tuchin,
  Nucl.\ Phys.\  A {\bf 717}, 249 (2003)
  [arXiv:nucl-th/0207037].
  
\bibitem{Adams:2003im}
  J.~Adams {\it et al.}  [STAR Collaboration],
  Phys.\ Rev.\ Lett.\  {\bf 91}, 072304 (2003)
  [arXiv:nucl-ex/0306024].

\bibitem{AGordon}
A.~Gordon (for the STAR Collaboration), Presentation at the 3rd Joint Meeting of APS Division of Nuclear Physics and Physical Society of Japan, Hawaii, October 13-17, 2009.

\bibitem{frag}
B.~A.~Kniehl, G.~Kramer and B.~Potter,
Nucl.\ Phys.\ B {\bf 597}, 337 (2001)
[arXiv:hep-ph/0011155].



\bibitem{Adams:2006uz}
  J.~Adams {\it et al.}  [STAR Collaboration],
  Phys.\ Rev.\ Lett.\  {\bf 97}, 152302 (2006)
  [arXiv:nucl-ex/0602011].


\bibitem{Braidot:2009ji}
  E.~Braidot  [STAR collaboration],
  Nucl.\ Phys.\  A {\bf 830}, 603C (2009)
  [arXiv:0907.3473 [nucl-ex]].
 \end{thebibliography}
\end{document}